\documentclass[showpacs,amsmath,amssymb, superscriptaddress]{revtex4}

\usepackage{amsmath}
\usepackage{amssymb}
\usepackage{amsfonts}

\usepackage{graphicx}

\usepackage{amsthm}
\DeclareMathOperator{\Tr}{Tr}
\newcommand{\U}{\it U}
\newcommand{\TTT}{{\rm T}}

\renewcommand{\Re}{{\rm Re}}

\makeatletter

\newcommand{\R}{{\mathbb R}}
\newcommand{\D}{{\cal D}}
\newcommand{\HH}{{\cal H}}

\newcommand{\LL}{{\cal L}^2}
\newcommand{\N}{{\cal N}}

\newcommand{\ket}[1]{|{#1}\rangle}
\newcommand{\kb}[1]{|{#1}\rangle\!\langle{#1}|}

\newcommand{\bra}[1]{\langle{#1}|}
\newcommand{\bkt}[2]{\langle{#1}|{#2}\rangle}

\newcommand{\hp}{\hat{p}}

\newtheorem{defin}{Definition}
\newtheorem{thm}{Theorem}
\newtheorem{lem}{Lemma}

\begin{document}
\title{Optimal Covariant Measurement of Momentum on a Half Line in Quantum Mechanics}
 \author{Yutaka Shikano}
 \email{shikano@th.phys.titech.ac.jp}
 \author{Akio Hosoya}
 \email{ahosoya@th.phys.titech.ac.jp}
 \affiliation{Department of Physics, Tokyo Institute of Technology, Tokyo, Japan}
 \date{\today}
\begin{abstract}
	We cannot perform the projective measurement of a momentum on a half line since it is not an observable. Nevertheless, 
	we would like to obtain some physical information of the momentum on a half line. 
	We define an optimality for measurement as minimizing the variance between an inferred outcome of the measured system 
	before a measuring process and a measurement outcome of the probe system after the measuring process, restricting our 
	attention to the covariant measurement studied by Holevo. 
	Extending the domain of the momentum operator on a half line by introducing a two dimensional Hilbert space 
	to be tensored, we make it self-adjoint and explicitly construct a model 
	Hamiltonian for the measured and probe systems. By taking the partial trace over the newly introduced 
	Hilbert space, the optimal covariant positive operator valued measure (POVM) of a momentum on a half line is 
	reproduced. We physically describe the measuring process to optimally evaluate the momentum of a particle on a half line.
\end{abstract}
\maketitle
\section{Introduction}
\begin{figure}
		\centering
		\includegraphics[width=7.8cm]{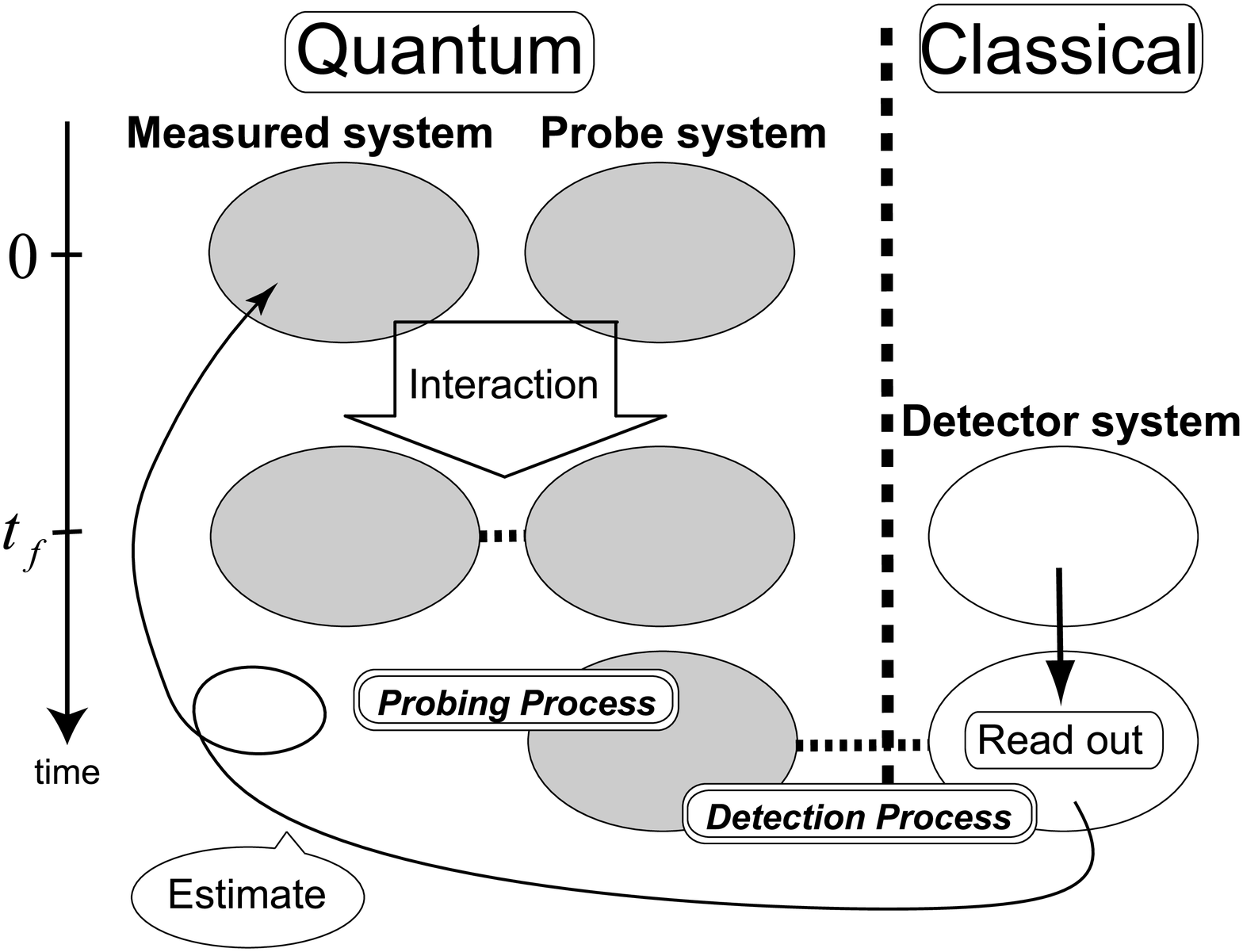}
		\caption{Scheme of measuring processes. We switch on the interaction between the measured 
		and probe systems in the first step to obtain the measurement outcome of the probe system in the 
		second step. We infer the observable of the measured system at $t = 0$ from the 
		outcome of the probe system at $t = t_{f}$ in the third step.}
		\label{measuring}
	\end{figure}

	Measurement in quantum mechanics is highly non-trivial as discussed in the mathematical foundation of quantum mechanics 
	initiated by von Neumann~\cite{neumann55}. He founded quantum mechanics on the Hilbert space and defined observables~\cite{comment3} 
	as self-adjoint operators to mathematically formulate the projection postulate in measuring processes 
	by the spectral theory in functional analysis. Davies and Lewis 
	constructed the framework of generalized measurement including the projective measurement by 
	introducing the concept of the {\textit{instrument}} and the positive operator valued 
	measure (POVM)~\cite{davies70}. Furthermore, by considering axioms 
	of measuring devices, Ozawa introduced the completely positive (CP) instrument~\cite{ozawa84} and showed that the state change by quantum 
	measuring processes can be described in terms of the Kraus operators~\cite{kraus71} and proposed a measuring 
	apparatus, i.e., a scheme of measurement consisting of a probing process described only by quantum mechanics and a 
	detection process described by the micro-macro coupling, which is to obtain classical information from quantum information
	~\cite{ozawa84,ozawa89}. To discuss measuring processes of the probe,
	we need to specify the Hilbert space corresponding to the probe system and an interaction Hamiltonian for the combined system of 
	the measured system and the probe system to calculate an evolution operator. 
	After the combined system is evolved in the measuring time, we acquire the measurement 
	outcome of the probe system and obtain the state of the measured system after the measuring 
	process taking the partial trace over the probe system. Note that we do not consider the detection process 
	to obtain observational data corresponding to a macroscopic experimental result~\cite{comment1}. All the above processes 
	are summarized in the book~\cite{busch91}, which is illustrated in Fig. \ref{measuring}. 
		
	For measuring processes, we shall consider an optimal measurement initiated by Helstrom~\cite{helstrom74}. 
	He defined an optimality of a measuring process to minimize the variance between an outcome of 
	a measured system {\textit{before}} the interaction and a measurement outcome of a probe system {\textit{after}} the 
	interaction. The optimal measurement sets upper limits to a POVM. In this paper, we explicitly construct a model 
	Hamiltonian which reproduces the optimal POVM in a special case, while a general method is not 
	available to construct a measurement model from a given POVM. 
	
	Let us recall that observables are defined as self-adjoint operators~\cite{comment2}. In a general case of symmetric operators,
	we cannot apply the projection postulate since symmetric operators cannot generally be decomposed 
	to real spectra, so that we have to consider generalized measurement~\cite{davies70}. 
	As an arch-typical example, we consider a half line system in quantum mechanics in this paper. 
	In a half line system, a momentum is not an observable as will be seen in Sec. \ref{chhalf}. Here, our present 
	proposal is to explicitly describe a generalized measuring process to optimally measure a momentum on a half line. 
	
	This paper is organized as follows. In Sec. \ref{chhalf}, we recapitulate the well-known property of a momentum operator 
	in quantum mechanics on a half line and will see that the momentum on a half line is not an observable. In Sec. \ref{chmeasurement}, 
	we define a covariant measurement and an optimality to measuring processes 
	and then reproduce the optimal covariant POVM introduced by Holevo~\cite{holevo78,holevo79,holevo82,holevo01}. In Sec. 
	\ref{chwholemodel}, we present a model Hamiltonian for the combined system of the measured and probe systems. 
	We calculate a POVM from the model Hamiltonian to reproduce the result corresponding to the optimal covariant POVM. Furthermore, 
	we present a physical description of our proposed measuring process. 
	In Sec. \ref{chhalfmodel}, we investigate the optimal covariant measurement 
	model of the momentum on a half line. To make the momentum operator on a half line self-adjoint, 
	we effectively extend the domain of this operator to the one of a whole line by tensoring a two dimensional Hilbert space. 
	We apply the optimal measurement model in Sec. \ref{chwholemodel} to the extended system. 
	Taking the partial trace over the extra Hilbert space, we obtain the optimal covariant 
	measurement model on a half line. 
	Section \ref{chconclusion} is devoted to the summary and discussion. 
\section{Quantum Mechanics on a Half Line} \label{chhalf}
	According to the functional analysis, on which the mathematical foundation of quantum mechanics~\cite{neumann55} is based, 
	an operator $\hat{A}$ is {\textit{symmetric}} if $\hat{A} = \hat{A}^{\dagger}$, where $\hat{A}^{\dagger}$ is the Hermite conjugate. 
	Further, a symmetric operator $\hat{A}$ is {\textit{self-adjoint}} if $\D (\hat{A}) = \D (\hat{A}^{\dagger})$, where $\D (\hat{A})$ 
	is the domain of the operator $\hat{A}$. In quantum mechanics, the observables are defined 
	as self-adjoint operators, which have real spectra~\cite{akhiezer81}. Symmetric operators, however, 
	do not necessarily have a real spectrum. We need to classify symmetric operators into self-adjoint operators, 
	essentially self-adjoint operators, self-adjoint extendable operators and non-self-adjoint 
	extendable operators~\cite{comment2}(for the definitions, see the book~\cite{akhiezer81}). A criterion is known as the deficiency 
	theorem (See Appendix \ref{deficiency}). 

	Let us specifically consider a quantum system on a half line $\R_{+} \equiv [0, \infty )$. 
	There have been many works concerning this problem since the beginning of quantum mechanics~\cite{rellich50, clark80, farhi90},
	e.g., the singular potential~\cite{case50,krall82,gordeyev97,fulop07}. 
	Recently, F\"{u}l\"{o}p {\it et al.} have studied boundary effects~\cite{fulop02,tsutsui03,fulop05} 
	and Twamley and Milburn have discussed a quantum measurement model on a half line by changing the coordinate $x \in \R_{+}$ to 
	$\log x \in \R$~\cite{twamley06}. 

	In the following consideration, we characterize the half line system as follows.
	Let us take a Hilbert space $\HH_{+} \equiv \LL (\R_{+})$ and a momentum operator $\hat{p}_{+}$ in $\HH_{+}$ defined by
	\begin{align}
		\hat{p}_{+} \psi (x) & = \frac{1}{i} \frac{d}{d x} \psi (x), \notag \\
		\D ({\hat{p}}_{+}) & = \left\{ \psi \in \HH_{+} ; \psi (0) = 0, \int^{\infty}_{0} 
		\left| \frac{d}{dx}\psi (x) \right|^2 dx < \infty \right\} 
		\label{momentumoperator}
	\end{align}
	in analogy to the standard momentum operator on a whole line. Throughout this paper, we take the unit $\hbar=1$.

	Then we can see that $\hat{p}_{+}$ is symmetric since 
	\begin{align}
		\bkt{\phi}{{\hat{p}}_{+} \psi} & = \frac{1}{i} \int^{\infty}_{0} \overline{\phi (x)} \frac{d}{dx} \psi (x) dx \notag \\
		& = \left[ \frac{1}{i} \overline{\phi (x)} \psi (x) \right]^{\infty}_{0} - \frac{1}{i}
		\int^{\infty}_{0} \frac{d}{dx} \overline{\phi (x)} \psi (x) dx \notag \\
		& = \int^{\infty}_{0} \overline{\frac{1}{i} \frac{d}{dx} \phi (x)} \psi (x) dx \notag \\
		& = \bkt{{\hat{p}}^{\dagger}_{+} \phi}{\psi} \label{p_dagger}, 
	\end{align}
	\begin{equation}
		\psi \in \D ({\hat{p}}_{+}) \qquad \phi \in \D ({\hat{p}}^{\dagger}_{+}),
	\end{equation}
	where ${\hat{p}}^{\dagger}_{+} = \frac{1}{i}\frac{d}{dx}$ with
	\begin{equation}
		\D ({\hat{p}}^{\dagger}_{+}) = \left\{ \psi \in \HH_{+} ; \int^{\infty}_{0} \left|\frac{d}{dx}\psi 
		(x) \right|^2 dx < \infty \right\} .
	\end{equation}
	Therefore we conclude that $(\hat{p}_{+}, \D (\hat{p}_{+})) \varsubsetneq ({\hat{p}}^{\dagger}_{+}, \D ({\hat{p}}^{\dagger}_{+}))$ 
	since $\D ({\hat{p}}_{+}) \neq \D ({\hat{p}}^{\dagger}_{+})$. So the momentum operator ${\hat{p}}_{+}$ on a half line 
	is symmetric but not self-adjoint, i.e., not an observable. 
\section{Review of Optimal Covariant Measurement} \label{chmeasurement}
	Let us consider a measuring process described by an interaction between a measured system and a probe system, the latter of which is 
	the part of the measuring apparatus as a whole. To establish the relationship between the measured and 
	probe systems, we consider the momentum space $\Omega = \R$ and a projective unitary 
	representation of the shift group of $\Omega$. Stone's theorem tells us that the unitary 
	representation is given by 
	\begin{equation}
		p \to V_{p} = e^{- i p \hat{x}},
	\end{equation}
	where $\hat{x}$ is the position operator.
	\begin{defin}
		A POVM $M(dp)$ is \textit{covariant} with respect to the 
		representation $p \to V_{p}$ if 
		\begin{equation}
			V_{p}^{\dagger} M(\Delta) V_{p} = M(\Delta_{-p}), \quad p \in \Omega
		\end{equation}
		for any $\Delta \in \mathcal{A}(\Omega)$, where 
		\begin{equation}
			\Delta_{p} = \{ p^{\prime} | p^{\prime} = p + p^{\prime \prime}, p^{\prime \prime} \in \Delta \}
			\label{range}
		\end{equation}
		is the image of the set $\Delta$ under the transformation $p$ and $\mathcal{A}(\Omega)$ is the Borel $\sigma$-field of $\Omega$.
	\end{defin}
	The covariant POVM has the property in the following form by using the Born formula~\cite{neumann55,holevo82},
	\begin{align}
		\Pr \{ \hat{p} \in \Delta_{p} \| \rho_{p + p^{\prime}_{0}} \} & = \Tr \rho_{p+p^{\prime}_0} M(\Delta_{p}) 
		= \Tr V_{-p} \rho_{p^{\prime}_0} V_{-p}^{\dagger} M(\Delta_{p}) 
		= \Tr \rho_{p^{\prime}_0} V_{-p}^{\dagger} M(\Delta_{p}) V_{-p} 
		= \Tr \rho_{p^{\prime}_0} M(\Delta) \notag \\
		& = \Pr \{ \hat{p} \in \Delta \| \rho_{p^{\prime}_{0}} \} .
	\end{align}
	That is, when the measured system is arbitrarily shifted, the measurement outcome is shifted by the same amount. 
	This idealized measurement is called a \textit{covariant measurement}. Realistic measuring 
	devices, however, satisfy this condition only locally as discussed by Hotta and Ozawa~\cite{hotta04}.

	By von Neumann's spectral theorem, any Hilbert space $\HH$ can be formally described as the direct integral 
	of a Hilbert space $\HH_{x}$,
	\begin{equation}
		\HH = \int \oplus \HH_x dx,
	\end{equation}
	so that any state vector $\psi \in \HH$ is described by the vector-valued function $\psi = [ \psi_x ]$ with $\psi_{x} \in \HH_x$ 
	introducing a convenient notation $[ \, \cdot \, ]$~\cite{holevo78,holevo82}. 
	There, a position operator $\hat{x}$ acts as multiplication operators
	\begin{equation}
		\hat{x} \psi = [ x \psi_x ] \label{holevoposition}
	\end{equation}
	in this notation. The same notation $[ \, \cdot \,]$ is used for an operator-valued function. 
	A kernel $[ K(x,x^{\prime}) ]$, where $K(x,x^{\prime})$ is a mapping from 
	$\HH_{x^{\prime}}$ to $\HH_{x}$ for all $x$ and $x^{\prime}$, defines an operator $\hat{K}$ on $\HH$. We can write 
	\begin{equation}
		\hat{K} \psi = \left[ K(x,x^{\prime}) \right] [ \psi_{x^{\prime}} ] = 
		\left[ \int K(x,x^{\prime}) \psi_{x^{\prime}} d x^{\prime} \right]. \label{holevooperator}
	\end{equation}
	Equations (\ref{holevoposition}) and (\ref{holevooperator}) can be rephrased by the bracket notation as 
	\begin{align}
		\hat{x} \ket{\psi} & = \int dx \ket{x} x \bkt{x}{\psi}, \\
		\hat{K} \ket{\psi} & = \int dx \int dx^{\prime} \ket{x} K (x, x^{\prime}) \bkt{x^{\prime}}{\psi}, 
	\end{align}
	respectively. Also we express the norm in $\HH_{x}$ as $\| \, \cdot \, \|_{x}$. 

	We are now in a position to explicitly describe the covariant POVM as follows. 
	\begin{thm}[Holevo~\cite{holevo78}]
		Any covariant POVM in $\HH$ has the form 
		\begin{equation} \label{covariantPOVM}
			M(dp) = \left[ K(x, x^{\prime}) e^{i(x - x^{\prime}) p} \frac{dp}{2 \pi} \right] ,
		\end{equation}
		where $[ K(x, x^{\prime}) ]$ is a positive definite kernel satisfying 
		$K(x, x) \equiv I_{x}$, the identity mapping from $\HH_{x}$ to itself.
		\label{holevo}
	\end{thm}
	In the above discussion, we have assumed that system and probe observables are isometric to obtain 
	(\ref{covariantPOVM}) as the POVM. The proof of Theorem \ref{holevo} is given in Appendix \ref{prf}. 

	Next we turn to a measuring process. First, we couple a measured system to a probe system. Second, 
	the combined system is evolved in time. Finally, we measure the probe observable. The sequence of processes enables us to 
	retrospectively evaluate the system observable at the starting time 
	by the measurement outcome of the probe observable at 
	the end time (See Fig. \ref{measuring}). So we define the optimal covariant measurement 
	as an optimal evaluation of the system observable by the outcome of the probe observable. 

	Let us assume that $W(p-P)$ is a deviation function, which expresses the variance between 
	the inferred "measurement" outcome $p$ of the system momentum before the interaction and 
	the measurement outcome $P$ of the probe momentum after the interaction, satisfying 
	\begin{equation}
		W(p) = - \int e^{i p x} \tilde{W} (dx),
	\end{equation}
	for an even finite measure $\tilde{W}(dx)$ on $\R$. Let us consider the condition to minimize the variance 
	\begin{equation}
		R_{p} \{ M \} = \int_{\Omega} W(p - P) \mu_{\rho} (dp),
		\label{deviation}
	\end{equation}
	where $\mu_{\rho} (dp) \equiv \Tr \rho M(dp)$ is the probability distribution for the pure state $\rho = \kb{\psi}$. 
	Because of covariance, we rewrite (\ref{deviation}) as 
	\begin{align}
		R_{0} \{ M \} & = \int_{\Omega} W(p) \mu_{\rho_0} (dp) \notag \\
		& = - \int \Phi_{\rho} (x) \tilde{W} (dx),
	\end{align}
	where 
	\begin{equation}
		\Phi_{\rho} (x) \equiv \int_{\Omega} e^{ixp} \bkt{\psi}{M(dp) \psi}
	\end{equation}
	is a characteristic function of $\mu_{\rho}(dp)$. 
	We get from Eq. (\ref{covariantPOVM})
	\begin{equation}
		\Phi_{\rho} (x) = \int \bkt{\psi_{\mu}}{K(\mu , \mu - x) \psi_{\mu - x}} d \mu .
	\end{equation}
	Since the integral converges by the Cauchy-Swartz inequality and the condition $K(x, x) = I_{x}$, 
	\begin{equation}
		\Re \Phi_{\rho} (x) \leq \Phi_{\ast} (x) \equiv \int \| \psi_{\mu} \|_{\mu} \| \psi_{\mu - x} \|_{\mu - x} d \mu ,
	\end{equation}
	so that
	\begin{align}
		R_{0} \{ M \} & \geq - \int \int \| \psi_{\mu} \|_{\mu} \| \psi_{\mu - x} \|_{\mu - x} d \mu \tilde{W} (dx) \notag \\
		& \equiv R_{0} \{ M_{0} \},
	\end{align}
	where
	\begin{equation}
		M_{0} (dp) = \left[ \frac{\psi_{x} \cdot \psi_{x^{\prime}}^{\dagger}}{\| \psi_{x} \|_{x} \| \psi_{x^{\prime}} \|_{x^{\prime}}} 
		e^{i (x - x^{\prime}) p} \frac{dp}{2 \pi} \right],
		\label{optimalcovariant}
	\end{equation}
	by transforming $\mu - x$ to $x^{\prime}$. Note that Eq. (\ref{optimalcovariant}) does not depend on the choice of the deviation 
	function $W( p - P )$ because of the covariance. It is curious to point out that this POVM corresponds to the optimal POVM under the 
	unbiased condition~\cite{hayashi00}. 
	In the case of the whole line system, the optimal covariant POVM (\ref{optimalcovariant}) in the bracket notation expresses 
	\begin{equation}
		M_{0} (dp) = \int_{\R} dx \int_{\R} dx^{\prime} \ket{x} e^{i (x - x^{\prime}) p} \frac{dp}{2 \pi} 
		\bra{x^{\prime}},
		\label{bracket}
	\end{equation}
	noting that the normalized term $\frac{\psi_{x} \cdot \psi_{x^{\prime}}^{\dagger}}{\| \psi_{x} \|_{x} \| 
	\psi_{x^{\prime}} \|_{x^{\prime}}}$ is the identity in the bracket notation. By using the Fourier transformation, 
	\begin{equation}
		\ket{p} = \frac{1}{\sqrt{2 \pi}} \int_{\R} dx e^{i p x} \ket{x},
	\end{equation}
	Eq. (\ref{bracket}) is transformed to the following equation:
	\begin{equation}
		M_{0} (dp) = \kb{p} dp,
	\end{equation}
	to obtain the projective measurement of a momentum on a whole line. 
	To summarize the above discussion, we obtain the optimal covariant POVM (\ref{optimalcovariant}) to minimize the 
	estimated variance between the system and probe observables~\cite{holevo78,holevo79}. We emphasize that Eq. (\ref{optimalcovariant}) 
	remains valid even when we change the domain of $x$.
\section{Optimal Measurement Model on a Whole Line} \label{chwholemodel}
	In the previous section, we have obtained the optimal covariant POVM. We are now going to explicitly construct a Hamiltonian 
	for a measurement model to realize the POVM. While it is straightforward to 
	calculate the POVM and the probability distribution of the system observable for a given Hamiltonian of a combined system, 
	it is not to find a Hamiltonian from a given POVM. In the two dimensional case, there is a way to construct 
	a model Hamiltonian from a given POVM~\cite{nielsen00}. Once the Hamiltonian for the combined system is found, 
	we can physically realize the given POVM in principle. In the infinite dimensional case, we heuristically explore 
	the optimal covariant POVM for the momentum in measuring processes in the following way. 
	In this section, we preparatively discuss measurement of the momentum of a particle on a whole line and then 
	apply the results to that on a half line in the next section. 
	To make our exposition shorter, we assume that 
	the wave functions $\{ \psi_{x} \}$ are normalized and the measure $\frac{dp}{2 \pi}$ is omitted 
	in Eq. (\ref{optimalcovariant}). Then Eq. (\ref{optimalcovariant}) is simply  
	\begin{equation}
		M_{0} = \left[ \psi_{x} \cdot \psi_{x^{\prime}}^{\dagger} e^{i (x - x^{\prime}) p} \right] .
		\label{newoptimal}
	\end{equation}

	Let us consider a model Hamiltonian, 
	\begin{align}
		{\hat{\HH}}_{com} & = \frac{1}{2m} {\hat{p}}^{2} + \frac{1}{2M} {\hat{P}}^{2} + g \hat{P} \hat{x} \delta (t) 
		+ \frac{m \omega^{2}}{2} {\hat{x}}^{2} \notag \\
		& \equiv \hat{\mathcal{H}}_{0} + g \hat{P} \hat{x} \delta (t), 
		\label{hamiltonian}
	\end{align}
	where a pair $(\hat{x}, \hat{p})$ are the position and the momentum operators of the measured system, a pair $(\hat{X}, \hat{P})$ 
	are those of the probe system and $\delta(t)$ is the Dirac $\delta$-function. 
	This Hamiltonian is modeled from the following consideration. We take the potential of the measured system 
	as a harmonic oscillator for simplicity and the probe system is assumed to be a free particle system. 
	Furthermore, the interaction is assumed to be instantaneous with a coupling constant $g$. 
	The interaction term $g \hat{x} \hat{P} \delta (t)$ is 
	chosen by the following reasoning. Because of the covariance, i.e., the measurement 
	value $\tilde{P}$ of the probe observable corresponds to the "measurement" value $\tilde{p}$ of 
	the system observable at a certain time, 
	we are led to an interaction of the momentum $\hat{P}$ of the probe system. Since the exponents in the optimal covariant POVM 
	(\ref{newoptimal}) has a quadratic form, a possible interaction term is either $g \hat{x} \hat{P}$ 
	or $g \hat{p} \hat{P}$. The latter is excluded because it does not influence the momentum of the measured system. 
	
	Furthermore, we assume that the measured system itself is weakly coupled to a bulk system at zero temperature. 
	We consider the measuring process from the time $t=0-$ to $t=t_f$. Then the evolution operator $\hat{\U}$ becomes 
	\begin{align} 
		\hat{\U} & = \TTT \; \exp \left( -i \int^{t_{f}}_{0-} {\hat{\HH}}_{com} dt \right) \notag \\
		& = \TTT \; \exp \left( -i \int^{t_{f}}_{\epsilon} {\hat{\HH}}_{0} dt \right) 
		\exp \left( -i \int^{\epsilon}_{- \epsilon} g \hat{P} \hat{x} \delta (t) dt \right) \notag \\
		& = \TTT \; \exp \left( -i \int^{t_{f}}_{\epsilon} {\hat{\HH}}_{0} dt \right) \exp \left( -i g \hat{P} \hat{x}(0) \right) ,
	\end{align}
	where $\epsilon$ is an infinitesimal positive parameter and $\TTT$ stands for the time-ordered product. 

	We construct the Kraus operator$[ {\hat{A}}_{x x^{\prime}} ]$ from the evolution operator as follows. Given the initial probe state 
	$\ket{\tilde{P}}$, an eigen state of the momentum $\hat{P}$ of the probe system, we see that 
	\begin{align}
		{\hat{A}}_{x x^{\prime}} & = \int \bra{P} \bra{x} \hat{\U} \ket{x^{\prime}} \ket{\tilde{P}} dP \notag \\
		& = \sum_{j} \bra{x} \TTT \; \exp \left( -i \int^{t_{f}}_{\epsilon} {\hat{\HH}}_{0} dt \right) \ket{j} 
		{\psi^{\dagger}}_{x^{\prime} , j} \exp \left( -ig \tilde{P} x (0) \right) \notag \\
		& \to \psi_{x} \cdot \psi^{\dagger}_{x^{\prime}} \exp \left( -ig \tilde{P} x (0) \right) \ \textrm{as} \; t_{f} \to \infty, 
		\label{Krausoperator}
	\end{align}
	where $\ket{P}$ is an eigen state of $\hat{P}$, $\psi_{x,j}$ is the wave function corresponding to the 
	$j$-th energy eigen state $\ket{j}$ and 
	$\psi = [ \psi_{x} ]$ is the ground state of the free Hamiltonian $\hat{\HH}_{0}$. In the last line of (\ref{Krausoperator}), 
	the ground state is picked up in the limit $t_f \to \infty$. Physically speaking, we measure the 
	probe observable after sufficient time passes. 
	Recall that the standard i$\epsilon$ prescription~\cite{abers73} implicitly assumes that the measured system itself 
	is weakly coupled to the bulk system at zero temperature. 
	Equation (\ref{Krausoperator}) is the matrix element of the Kraus operator $[ {\hat{A}}_{x x^{\prime}} ]$. 

	From the Kraus operator, we calculate the POVM as 
	\begin{align}
		M & = \left[ \int {\hat{A}}^{\dagger}_{x^{\prime} x^{\prime \prime}} {\hat{A}}_{x x^{\prime \prime}} d x^{\prime \prime} 
		\right] \label{kraus} \\
		& = \left[ \psi^{\dagger}_{x^{\prime}} \cdot \psi_{x} \exp \left( -ig \tilde{P} 
		\{ x (0) - x^{\prime} (0) \} \right) \right] .
		\label{measurementmodel}
	\end{align}
	We identify $g \tilde{P}$ with the measurement outcome $P$ itself of the probe observable to   
	reproduce the optimal covariant POVM (\ref{newoptimal}). 

	\begin{figure}
		\centering
		\includegraphics[width=8cm]{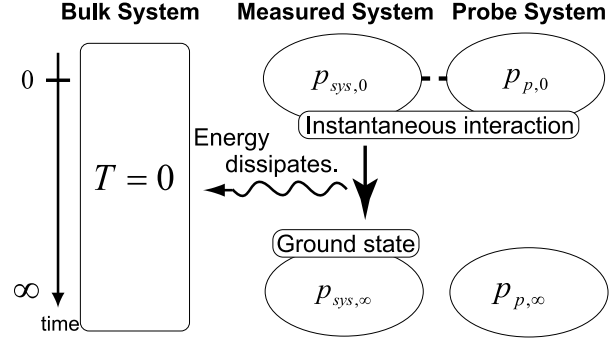}
		\caption{An optimal covariant measurement model. By the instantaneous interaction between the measured 
		and probe systems, the measured system is entangled with the probe system. On the other hand, the measured 
		system is coupled with the bulk system at zero temperature to dissipate the energy of the measured system. Thus we 
		optimally evaluate the system observable at $t = 0$ inferred from the outcome of 
		the probe system at $t = \infty$ by the momentum conservation law.}
		\label{model}
	\end{figure}
	Now, we physically describe how we optimally infer the momentum of the measured system 
	just before the measuring process. 
	First, we instantaneously couple the measured system to the probe system. Second, we keep the measured system in contact with 
	the bulk system at zero temperature and wait for a sufficiently long time. 
	Since the energy of the measured system is dissipated to the bulk system, the state of the measured system settles 
	down to the ground state. 
	If we let the energy of the ground state zero, i.e., $\omega \to 0$ of the interaction Hamiltonian (\ref{hamiltonian}), 
	the momentum of the measured system $p_{sys, \infty}$ becomes zero at $t_f = \infty$. According to the momentum 
	conservation law, we obtain 
	\begin{equation}
		p_{sys, 0} + p_{p, 0} = p_{sys, \infty} + p_{p, \infty} = p_{p, \infty}, 
		\label{conservation}
	\end{equation}
	where $p_{sys, t}$ and $p_{p, t}$ are the momenta of the measured system and the probe system at a time $t$.
	Since we can control the probe system, we can precisely infer the "measurement" value $p_{sys, 0}$ of the momentum 
	of the measured system at the beginning of the measuring process from the measurement outcome $p_{p, \infty}$, 
	which we measure in the probe system at $t_f = \infty$ (See Fig. \ref{model}). 
	If $\omega$ of the Hamiltonian (\ref{hamiltonian}) were finite, 
	the variance of the momentum of the measured system would remain finite due to the zero point oscillation 
	and Eq. (\ref{conservation}) would be modified. 

	Although we have assumed that the potential of the measured system is given by the harmonic oscillator, 
	the potential could actually be any convex function since the i$\epsilon$ prescription picks up the ground state 
	at $t_f \to \infty$. 
\section{Optimal Measurement Model on a Half Line} \label{chhalfmodel}
\begin{figure}
		\centering
		\includegraphics[width=8cm]{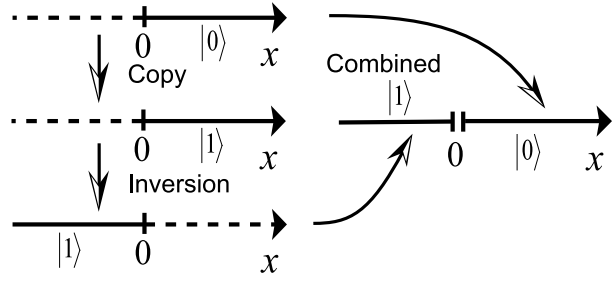}
		\caption{A Naimark extension. An auxiliary two dimensional Hilbert space $\HH_{2}$ is tensored to the  
		Hilbert space $\HH_{+}$ to prepare the 
		two (original and copied) Hilbert spaces. Then we spatially invert the copied Hilbert space around 
		the zero point. Finally, we combine the original and inverted Hilbert spaces to 
		obtain the extended Hilbert space, $\HH = \HH_{+} 
		\otimes \HH_{2} = \HH_{+} \oplus \HH_{-}$.}
		\label{naimark}
	\end{figure}

	Let us apply the optimal measurement model to the half line system. We have already seen that the 
	momentum operator (\ref{momentumoperator}) is not self-adjoint. First, we extend the domain of 
	$\hat{p}_{+}$ {\textit{\'{a} la}} Naimark so that the extended operator $\hat{p}$ is self-adjoint. 
	The extended Hilbert space is 
	\begin{equation}
		\HH = \HH_{+} \otimes \HH_{2},
	\end{equation}
	where $\HH \equiv \LL (\R)$, $\HH_{+} \equiv \LL (\R_{+})$ and $\HH_{2}$ is the two dimensional Hilbert space of the 
	two level system with the orthonormal bases $\ket{0}$ and $\ket{1}$. We choose the form of the extended momentum operator as 
	\begin{equation}
		\hp = \hp_{+} \otimes \kb{0} - \hp_{+} \otimes \kb{1}. 
		\label{exam}
	\end{equation}
	By the unitary transformation $\Pi_{1}$, which is the space inversion around the zero point only for the spin state $\ket{1}$, 
	the Hilbert space $\HH$ is unitarily equivalent to 
	\begin{equation}
		\HH = \HH_{+} \otimes \ket{0} + \HH_{-} \otimes \ket{1} = \HH_{+} \oplus \HH_{-}, 
	\end{equation}
	where $\HH_{-} \equiv \LL (\R_{-})$ and $\R_{-} \equiv (- \infty, 0]$. 
	Then we transform the extended momentum operator (\ref{exam}) by $\Pi_{1}$ as 
	\begin{equation}
		\Pi_{1} \hp \Pi^{\dagger}_{1} = \hp_{+} \otimes \kb{0} + \hp_{-} \otimes \kb{1},
		\label{extendedmomentum}
	\end{equation}
	where $\hp_{+}$ and $\hp_{-}$ are momentum operators, which have the following domains 
	\begin{align}
		\D (\hp_{+}) & = \left\{ \psi \in \HH_{+} \ ; \ \psi (0) = 0, \int_{0}^{\infty} \left| \frac{d}{dx} 
		\psi (x) \right|^{2} < \infty \right\} \notag \\
		\D (\hp_{-}) & = \left\{ \psi \in \HH_{-} \ ; \ \psi (0) = 0, \int_{- \infty}^{0} \left| 
		\frac{d}{dx} \psi (x) \right|^{2} < \infty \right\}, 
	\end{align}
	respectively. Then the extended operator $\hp$ is self-adjoint extendable since the domain is the Hilbert space for 
	the whole line system. For a more precise argument, see Appendix \ref{deficiency}, where the choice of a 
	boundary condition $\psi (0) = 0$ is also justified. These operations are exhibited in Fig. \ref{naimark}. 
	It is curious to point out that this operator $\hp$ is $\cal{PT}$ symmetric noting 
	that the spin states $\ket{0}$ and $\ket{1}$ are interchanged by the time reversal $\cal{T}$ and the momentum operators 
	$\hp_{+}$ and $\hp_{-}$ by the parity inversion and the time reversal $\cal{PT}$. 
	We can see that the spectrum of $\hp$ is real also from this reasoning~\cite{bender07}. 

	We adopt the form of the model Hamiltonian (\ref{hamiltonian}) with $\hp$ being replaced by the right hand side of 
	(\ref{extendedmomentum}) and $x \in \R$, so that all the operators in the Hamiltonian (\ref{hamiltonian}) are 
	self-adjoint to construct the optimal covariant 
	measurement in the same way as described in Sec. \ref{chwholemodel}. 
	We, then, calculate the Kraus operator from the model Hamiltonian by using the i$\epsilon$ prescription. Since we 
	have chosen $\psi (0) = 0$, we end up with the ground state with odd parity with the energy $\frac{3}{2} \omega$. 
	The Kraus operator is then 
	\begin{align} 
		\Pi_{1} [ {\hat{A}}_{x x^{\prime}} ] \Pi^{\dagger}_{1} = \left[ \psi_{x_{+}} \cdot \psi^{\dagger}_{x^{\prime}_{+}} 
		\exp \left( -ig P_{+} x_{+} (0) \right) \right] 
		\otimes \kb{0} + \left[ \psi_{x_{-}} \cdot \psi^{\dagger}_{x^{\prime}_{-}} \exp 
		\left( -ig P_{-} x_{-} (0) \right) 
		\right] \otimes \kb{1} .
		\label{halflinekraus}
	\end{align}
	From Eq. (\ref{kraus}), the Kraus operator (\ref{halflinekraus}) gives the following POVM, 
	\begin{equation} 
		\Pi_{1} M_{0} \Pi^{\dagger}_{1} = \left[ \psi_{x_{+}} \cdot \psi_{x^{\prime}_{+}}^{\dagger} 
		e^{i (x_{+} - x^{\prime}_{+}) p_{+}} \right] 
		\otimes \kb{0} + \left[ \psi_{x_{-}} \cdot \psi_{x^{\prime}_{-}}^{\dagger} e^{i (x_{-} - x^{\prime}_{-}) 
		p_{-}} \right] \otimes \kb{1}.
		\label{halflineoptimal}
	\end{equation}
	By taking the partial trace over $\HH_{2}$, we obtain the reduced POVM 
	\begin{align}
		{\tilde{M}}_{0} & \equiv \Tr_{2} M_{0} \notag \\
		& = \left[ \psi_{x_{+}} \cdot \psi^{\dagger}_{{x^{\prime}_{+}}} e^{i (x_{+} - x^{\prime}_{+}) p_{+}} \right] ,
		\label{mm}
	\end{align}
	up to a normalization constant. Here in Eq. (\ref{mm}), we have transformed (\ref{halflineoptimal}) back to $M_{0}$ 
	by the unitary operator $\Pi_{1}$ and reproduced the optimal covariant POVM (\ref{newoptimal}) restricted to positive 
	parameters $x$ and $x^{\prime}$. 
	
	Finally, we calculate the probability distribution of the momentum on a half line in the optimal case. As an example, let us 
	assume the pure state $\rho = \left[ \phi_{x_{+}} \cdot \phi^{\dagger}_{x^{\prime}_{+}} \right]$, which is a plane wave 
	with a momentum $p_{true}$, 
	\begin{equation}
		\phi_{x_{+}} = A e^{i p_{true} x_{+}},
		\label{plane}
	\end{equation}
	for the measured system before the measuring process. We assume that the state (\ref{plane}) is properly localized to be 
	an element of the Hilbert space $\HH_{+}$. 
	The state (\ref{plane}), $\left[ \phi_{x_{+}} \right]$, is relaxed by the measuring process to the ground state $\psi_{x_{+}} \in \HH_{+}$ given by  
	\begin{equation}
		\psi_{x_{+}} = 2 \left( \frac{ (m \omega)^3}{\pi} \right)^{\frac{1}{4}} x_{+} 
		\exp \left( - \frac{m \omega}{2} x^{2}_{+} \right). 
	\end{equation}
	Then we obtain the probability distribution of the momentum as  
	\begin{align} 
		\Tr (\rho {\tilde{M}}_{0}) & = \Tr \left( \left[ \phi_{x^{\prime \prime}_{+}} \cdot \phi^{\dagger}_{x^{\prime}_{+}} \right] 
		\left[ \psi_{x_{+}} \cdot \psi^{\dagger}_{{x^{\prime \prime}_{+}}} e^{i (x_{+} - x^{\prime \prime}_{+}) p} \right] 
		\right) \notag \\
		& = \int \int \phi_{x^{\prime \prime}_{+}} \cdot \phi^{\dagger}_{x_{+}} \cdot \psi_{x_{+}} \cdot \psi^{\dagger}_{{x^{\prime 
		\prime}_{+}}} e^{i (x_{+} - x^{\prime \prime}_{+}) p} dx dx^{\prime \prime} \notag \\
		& = 16 \sqrt{\frac{\pi}{(m \omega)^{3}}} |A|^2  \left( p - p_{true} \right)^{2} 
		\exp \left( - \frac{1}{m \omega} \left( p - p_{true} \right)^{2} \right),
	\end{align}
	which has two peaks at $p = p_{true} \pm \sqrt{m \omega}$ and vanishes at $ p=p_{true} $. 
	If we take $\omega \to 0$, i.e., the measured system is a 
	free particle system, we can precisely evaluate the momentum of the plane wave since we obtain 
	$\Tr (\rho {\tilde{M}}_{0}) = \delta (p -p_{true})$. Otherwise there remains uncertainty by quantum zero point oscillation 
	and the momentum with the maximum probability deviates by $\sqrt{m \omega}$ from the precise momentum $p_{true}$. 
	When the potential of the measured system is a general convex function, the probability distribution for 
	the momentum becomes the modulus square of the Fourier transformation of the odd parity ground state wave function.
	
	To summarize this section, we have obtained the optimal covariant POVM on a half line, which enables us to explicitly 
	construct the measuring process of the momentum on a half line. 
\section{Summary and Discussion} \label{chconclusion}
	We have considered the optimal covariant measurement of momenta on a half line. Since the momentum operator $\hp_{+} = 
	\frac{1}{i} \frac{d}{dx}$ on a half line is not self-adjoint, i.e., not an observable. 
	By applying the Naimark extension, the measured system is extended to the whole line and the momentum operator 
	on the extended system becomes self-adjoint. Then we have discussed the optimal covariant measurement model 
	on the extended system. By applying Holevo's works~\cite{holevo78,holevo79,holevo82,holevo01}, we have obtained 
	the optimal covariant POVM in the optimal sense to minimize the variance between the "measurement" outcome of 
	the measured system before the interaction and the measurement outcome of the probe system after the interaction. 
	To realize physical systems, we have explicitly constructed the model Hamiltonian for the measured and probe systems and 
	coupled the measured system to the bulk system at zero temperature for infinitely long time. 
	We have shown that the optimal covariant POVM coincides with the calculated POVM from the model Hamiltonian. 
	As a result, we have presented the optimal covariant measurement model. Then we have physically explained the optimal 
	covariant measuring process. By taking the partial trace over the auxiliary Hilbert space $\HH_{2}$, 
	we have described the optimal covariant measurement model 
	for the momentum on a half line and calculated the optimal probability distribution of the momentum on a half line 
	in a special case. 
	
	The following points remain to be clarified. First, we have only discussed the covariant case. 
	Peres and Scudo, however, pointed out that the covariant measurement 
	may not be optimal and mentioned counterexamples in quantum phase measurement~\cite{peres02}. 
	We have to check whether the optimality for any measurement is the optimal covariant measurement in our setup or not. 
	Second, Ozawa have recently constructed a new Heisenberg uncertainty 
	principle~\cite{ozawa03,ozawa04}. The inequality expresses a quantum limit of 
	measuring processes. It will be interesting to examine Ozawa's inequality in our framework. 
	Third, there is an analogy between a momentum operator on a half line and a time or time-of-arrival operator since a energy has a lower bound. 
	However, there has been a long debate about mathematical formulations and physical meanings of a time or time-of-arrival 
	operator (for example \cite{holevo82, aharonov61, baute00, arai08}). It will be interesting to show physical meanings of this operator motivated 
	by our framework. 
	Finally, we have presented the model Hamiltonian (\ref{hamiltonian}) to physically realize the optimal covariant POVM
	(\ref{optimalcovariant}). We do not know a general method to construct a Hamiltonian from an arbitrary POVM. Our analysis may be 
	a clue to the general method to solve the inverse problem. 
	Furthermore, to experimentally demonstrate the measurement model, experimental 
	setups remain to be considered for our proposed model Hamiltonian. 
\section*{Acknowledgment}
	We would like to thank Mr. Yasumichi Matsuzawa, Mr. Takahiro Sagawa and Professor Shogo Tanimura for useful comments 
	and Professor Masanao Ozawa for his kind suggestion. 
\appendix
\section{Proof of Theorem \ref{holevo}} \label{prf}
	A measure $\nu$ is called \textit{invariant} on $\mathcal{A}(\Omega)$ if and only if for any $p \in \Omega$ there exists 
	a measure $\nu$ such that 
	\begin{equation}
		\nu (\Delta_{p}) = \nu (\Delta ),
	\end{equation}
	where $\Delta$ is an element of $\mathcal{A}(\Omega)$ and $\Delta_{p}$ is defined in (\ref{range}).

	The following lemma is useful.
	\begin{lem} \label{le}
		Let $M(dp)$ be a covariant POVM with respect to a projective unitary representation $p \to V_{p}$ of 
		the parametric group $G$ of transformations of the set $\Omega$. For any density operator $\rho$ on 
		the Hilbert space of the representation and for any Borel set $\Delta \in \mathcal{A}(\Omega)$
		\begin{equation}
			\int_{\Omega} \Tr V_{p} \rho V^{\dagger}_{p} M(\Delta) \mu (dp) = \nu (\Delta)
			\label{lem}
		\end{equation}
		where $\mu (dp)$ is the proper Lebesgue measure and the extent $\Omega$ of the integral 
		is a space of parameters in $G$ and 
		$\nu$ is an invariant measure. 
	\end{lem}
	\proof{
	Define 
	\begin{equation}
		\mathbf{1}_{\Delta} (p) = \biggl\{ \begin{array}{ll}
			1 & \textrm{on} \; p \in \Delta \\
			0 & \textrm{on} \; p \notin \Delta
		\end{array} .
	\end{equation}
	Then we can write the left hand side of (\ref{lem}) as
	\begin{equation}
		\int_{\Omega} \Tr V_{p} \rho V^{\dagger}_{p} M(\Delta) \mu (dp) 
		= \int_{\Omega} \Tr \rho M(\Delta_{-p}) \mu (dp) 
		= \int_{\Omega} \int_{\Omega} \mathbf{1}_{\Delta} (p + p_{0}) \mu_{\rho} (d p_{0}) \mu (dp), 
	\end{equation}
	where $\mu_{\rho} (dp_{0}) \equiv \Tr \rho M (dp_{0})$ is the probability distribution of the momentum for the state $\rho$ 
	noting that the second integral on the rightmost side is over the whole momentum space $\Omega$. 
	We see that 
	\begin{equation}
		\int_{\Omega} \mu_{\rho} (d p_{0}) \int_{\Omega} \mathbf{1}_{\Delta} (p + p_{0}) \mu (dp) = \nu (\Delta).
	\end{equation}}
	
	The following is the proof of Theorem \ref{holevo}.
	\proof{
	Let us assume $\rho =  \ket{\psi} \bra{\psi}$ without loss of generality. Then we see that 
	\begin{equation}
		\int_{\Omega} V_{p} \rho V^{\dagger}_{p} \frac{dp}{2 \pi}
		 = \left[ \int_{\Omega} e^{i p x} \psi_{x} \cdot \psi^{\dagger}_{x^{\prime}} e^{-i p 
		 x^{\prime}} \frac{dp}{2 \pi} \right]
		 = \left[ \delta (x - x^{\prime}) \psi_{x} \cdot \psi^{\dagger}_{x} \right] . \label{pl}
	\end{equation}
	Noting that the operator $M(\Delta)$ is defined by the kernel $M_{\Delta} (x, x^{\prime})$, we obtain from Lemma \ref{le} 
	and Eq. (\ref{pl})
	\begin{equation}
		\int_G \psi^{\dagger}_{x} M_{\Delta} (x,x) \psi_{x} dx = \frac{\mathrm{mes} \Delta}{2 \pi},
	\end{equation}
	where $G$ is the parametric group and $\mathrm{mes} \Delta$ denotes the Lebesgue measure 
	of $\Delta$. Since $[ \psi_{x} ]$ is arbitrary, we see that 
	$M_{\Delta} (x, x) = \frac{1}{2 \pi} \mathrm{mes} \Delta \cdot I_{x}$ noting that $I_x$ is the identity mapping from 
	$\HH_x$ to itself. From the positive definiteness of 
	$M_{\Delta} (x, x^{\prime})$, we can derive that
	\begin{equation}
		| \psi^{\dagger}_{x} M_{\Delta} (x,x^{\prime}) \psi_{x^{\prime}} | 
		\leq \sqrt{\psi^{\dagger}_{x} M_{\Delta} (x,x) \psi_{x}} 
		\sqrt{\psi^{\dagger}_{x^{\prime}} M_{\Delta} (x^{\prime},x^{\prime}) \psi_{x^{\prime}}}
		= \| \psi_{x} \|_{x} \| \psi_{x^{\prime}} \|_{x^{\prime}} \frac{\mathrm{mes} \Delta}{2 \pi}
	\end{equation}
	using the Cauchy-Schwartz inequality. 
	Therefore, the measure $M_{\Delta} (x, x^{\prime})$ is absolutely continuous with respect to the Lebesgue measure, so that 
	we can express 
	\begin{equation}
		M_{\Delta} (x, x^{\prime}) = \frac{1}{2 \pi} \int_{\Delta} K_{p} (x, x^{\prime}) dp,
	\end{equation}
	with $K_{p}(x, x^{\prime})$ being some positive definite density satisfying $K_{p}(x,x) = I_{x}$. 
	From the covariance properties, it follows that
	\begin{equation}
		K_{p} (x, x^{\prime}) = e^{i (x-x^{\prime}) p} K_{0} (x,x^{\prime}).
	\end{equation}
	Putting $K_{0} (x,x^{\prime}) = K(x,x^{\prime})$, we get (\ref{covariantPOVM}) in Theorem \ref{holevo}.}
\section{Deficiency Theorem} \label{deficiency}
	We refer the reader to the book~\cite{reed75} and the paper~\cite{bonneau01} for details. We shall 
	give a criterion for closed symmetric operators to be self-adjoint operators. 

	Let us assume that $(\hat{A}, \D (\hat{A}))$ is densely defined, symmetric and closed. 
	One defines the deficiency subspaces $\N_{\pm}$ by, for a fixed $\gamma > 0$, 
	\begin{gather}
		\N_{+} = \{ \psi \in \D (\hat{A}^{\dagger}) \, ; \ \hat{A}^{\dagger} \psi = i \gamma \psi \} \\
		\N_{-} = \{ \psi \in \D (\hat{A}^{\dagger}) \, ; \ \hat{A}^{\dagger} \psi = - i \gamma \psi \}
	\end{gather}
	of respective dimensions $n_{+}$ and $n_{-}$, which are called the deficiency indices of the operator $\hat{A}$ and 
	denoted by a pair $(n_{+}, n_{-})$. The following theorem holds. 
	\begin{thm}[Deficiency theorem]
		For any closed symmetric operator $\hat{A}$ with deficiency indices $(n_{+}, n_{-})$, there are three 
		possibilities: 
		\begin{enumerate}
			\item $\hat{A}$ is self-adjoint if and only if $n_{+} = n_{-} = 0$.
			\item $\hat{A}$ has self-adjoint extensions if and only if $n_{+} = n_{-}$. There exists one-to-one correspondence 
				between self-adjoint extension of $\hat{A}$ and unitary maps from $\N_{+}$ to $\N_{-}$.
			\item If $n_{+} \neq n_{-}$, $\hat{A}$ has no self-adjoint extension.
		\end{enumerate}
	\end{thm}
	This theorem is firstly discussed by Weyl~\cite{weyl10} and generalized by 
	von Neumann~\cite{neumann29}. 
	
	Let us apply this theorem to the momentum operator (\ref{momentumoperator}) on a half line. 
	First, we solve the differential equations,  
	\begin{equation}
		\hat{p}_{+} \psi_{\pm} (x) = - i \frac{d}{dx} \psi_{\pm} (x) = \pm i \gamma \psi_{\pm} (x) ,
	\end{equation}
	where $\gamma$ is real and positive to obtain 
	\begin{equation}
		\psi_{\pm} (x) \sim e^{\mp \gamma x}.
	\end{equation}
	Because of $\psi \in \LL (\R_{+})$, only $\psi_{+} (x)$ is allowed. Therefore, we obtain 
	the deficiency indices $(1,0)$ and conclude, by the deficiency theorem, $\hat{p}_{+}$ has no self-adjoint extension. 
	
	As another example, we show that the extended momentum operator (\ref{exam}) is self-adjoint extendable. 
	We obtain the deficiency indices $(0,1)$ of $-\hp_{+}$ in the same way. So the deficiency indices of the 
	extended momentum operator (\ref{exam}) are $(1,1)$ and the operator is self-adjoint extendable by the deficiency theorem. 
	Since the self-adjoint extension is parametrized by $U(1)$, $\psi (0+) = e^{i \theta} \psi (0-)$ where $\theta \in \R$, 
	we have a freedom to choose the boundary conditions at the 
	origin by that amount. The boundary condition $\psi (0) = 0$ chosen in the main text, which comes from the physical requirement 
	to the half line system, is mathematically legitimate in the extended system because it is a special case of the $U(1)$ variety. 


\begin{thebibliography}{99}
\bibitem{neumann55}
	J. von Neumann,
	\textit{Mathematische Grundlagen der Quantumechanik} 
	(Springer, Berlin, 1932),
	[ \textit{Mathematical foundations of quantum mechanics} 
	(Princeton University Press, Princeton, 1955). ]
\bibitem{comment3}
	The "observable" is a technical term. We use this term as a self-adjoint operator {\textit{\'{a} la}} von Neumann by the Kato-Rellich theorem 
	while the use of this term might be controversial.
\bibitem{davies70}
	E. B. Davies and J. T. Lewis, 
	Commun. Math. Phys. 
	\textbf{17}, 239-260 (1970).
\bibitem{ozawa84}
	M. Ozawa, 
	J. Math. Phys. 
	\textbf{25}, 79-87 (1984).
\bibitem{kraus71}
	K. Kraus, 
	Ann. Phys. 
	\textbf{64}, 311-335 (1971).
\bibitem{ozawa89}
	M. Ozawa, 
	in \textit{Squeezed and Nonclassical Light}, 
	edited by P. Tombesi and E. R. Pike, 
	(Plenum, New York, 1989), 
	pp. 263- 286.
\bibitem{comment1}
	We often call the detection process a magnification process, which is how to observe a pointed value of 
	measuring devices, e.g., physical processes in a photomultiplier. This process is discussed 
	by many people, e.g., see Ojima~\cite{ojima07}. 
\bibitem{busch91}
	P. Busch, P. Mittelstaedt and P. J. Lahti, 
	\textit{Quantum Theory of Measurement} 
	(Springer-Verlag, Berlin, 1991).
\bibitem{helstrom74}
	C. W. Helstrom, 
	Int. J. Theor. Phys. 
	\textbf{11}, 357-378 (1974).
\bibitem{comment2}
	Many physicists do not classify symmetric operators into self-adjoint 
	operators and often call a symmetric operator Hermitian and identify 
	a Hermitian operator with an observable without checking a domain of a operator (See, e.g.,
	~\cite{schiff65,landau77}).
\bibitem{holevo78}
	A. S. Holevo,
	Rep. Math. Phys.
	\textbf{13}, 379-399 (1978).
\bibitem{holevo79}
	A. S. Holevo,
	Rep. Math. Phys.
	\textbf{16}, 385-400 (1979).
\bibitem{holevo82}
	A. S. Holevo,
	\textit{Probabilistic and statistical aspects of quantum theory}
	(North-Holland, Amsterdam, 1982).
\bibitem{holevo01}
	A. S. Holevo,
	\textit{Statistical Structure of Quantum Theory}
	(Springer, Berlin, 2001).
\bibitem{akhiezer81}
	N. I. Akhiezer and I. M. Glazman,
	\textit{Theory of Linear Operators in Hilbert Space} 
	(Dover, New York, 1993).
\bibitem{rellich50}
	F. Rellich, 
	Math. Ann. 
	\textbf{122}, 343-368 (1950).
\bibitem{clark80}
	T. E. Clark, R. Menioff and D. H. Sharp, 
	Phys. Rev. D 
	\textbf{22}, 3012-3016 (1980).
\bibitem{farhi90}
	E. Farhi and S. Gutmann, 
	Int. J. Mod. Phys. A
	\textbf{5}, 3029-3051 (1990).
\bibitem{case50}
	K. M. Case, 
	Phys. Rev. 
	\textbf{80}, 797-806 (1950).
\bibitem{krall82}
	A. M. Krall,
	J. Diff. Eq. 
	\textbf{45}, 128-138 (1982).
\bibitem{gordeyev97}
	A. N. Gordeyev and S. C. Chhajlany, 
	J. Phys. A 
	\textbf{30}, 6893-6909 (1997).
\bibitem{fulop07}
	T. F\"{u}l\"{o}p, 
	arXiv:0708.0866.
\bibitem{fulop02}
	T. F\"{u}l\"{o}p, T. Cheon and I. Tsutsui, 
	Phys. Rev. A 
	\textbf{66}, 052102 (2002).
\bibitem{tsutsui03}
	I. Tsutsui, T. F\"{u}l\"{o}p and T. Cheon, 
	J. Phys. A 
	\textbf{36}, 275-287 (2003).
\bibitem{fulop05}
	T. F\"{u}l\"{o}p,
	Ph.D. thesis, University of Tokyo, 2005.
\bibitem{twamley06}
	J. Twamley and G. J. Milburn,
	New J. of Phys.
	\textbf{8}, 328 (2006).
\bibitem{hotta04}
	M. Hotta and M. Ozawa,
	Phys. Rev. A
	\textbf{70}, 022327 (2004).
\bibitem{hayashi00}
	M. Hayashi and F. Sakaguchi,
	J. Phys. A
	\textbf{33}, 7793 (2000). 
\bibitem{nielsen00}
	M. A. Nielsen and I. L. Chuang, 
	\textit{Quantum Computation and Quantum Information} 
	(Cambridge University Press, Cambridge, 2000).
\bibitem{abers73}
	E. S. Abers and B. W. Lee, 
	Phys. Rep. 
	\textbf{9}, 1-141 (1973).
\bibitem{bender07}
	C. M. Bender, 
	Rep. Prog. Phys. 
	\textbf{70}, 947-1018 (2007).
\bibitem{peres02}
	A. Peres and P. Scudo, 
	J. Mod. Opt. 
	\textbf{49}, 1235-1243 (2002).
\bibitem{ozawa03}
	M. Ozawa, 
	Phys. Rev. A 
	\textbf{67}, 042105 (2003).
\bibitem{ozawa04}
	M. Ozawa, 
	Ann. Phys. 
	\textbf{311}, 350-416 (2004).
\bibitem{aharonov61}
	Y. Aharonov and D. Bohm, 
	Phys. Rev. 
	\textbf{122}, 1649 (1961).
\bibitem{baute00}
	A. D. Baute, I. L. Egusquiza, J. G. Muga and R. Sala Mayato, 
	Phys. Rev. A 
	\textbf{61}, 052111 (2000).
\bibitem{arai08}
	A. Arai and Y. Matsuzawa, 
	mp-arc:08-24 (2008). 
\bibitem{reed75}
	M. Reed and B. Simon,
	\textit{Methods of Modern Mathematical Physics II, Fourier Analysis, Self-Adjointness}
	(Academic Press, New York, 1975).
\bibitem{bonneau01}
	G. Bonneau, J. Faraut and G. Valent, 
	Am. J. Phys. 
	\textbf{69}, 322-331 (2001).
\bibitem{weyl10}
	H. Weyl, 
	Math. Ann. 
	\textbf{68}, 220-269 (1910).
\bibitem{neumann29}
	J. von Neumann, 
	Math. Ann. 
	\textbf{102}, 49-131 (1929).
\bibitem{ojima07}
	I. Ojima, 
	arXiv:0705.2945.
\bibitem{schiff65}
	L. I. Schiff, 
	\textit{Quantum Mechanics}
	(McGraw-Hill, 1965) Third edition.
\bibitem{landau77}
	L. D. Landau and E. M. Lifschitz,
	\textit{Quantum Mechanics Non-Relativistic Theory}
	(Pergamon Press, 1977) Third edition.
\end{thebibliography}
\end{document}